\documentclass[12pt,preprint]{aastex}

\shorttitle{Interaction of EUV Wave with Active Region Loops}
\shortauthors{Yang et al.}

\begin{document}

\title{$\emph{SDO}$/AIA and $\emph{Hinode}$/EIS Observations of Interaction Between an EUV Wave and Active Region Loops}

\author{Liheng Yang\altaffilmark{1}, Jun Zhang\altaffilmark{1}, Wei Liu\altaffilmark{2}, Ting Li\altaffilmark{1} and Yuandeng Shen\altaffilmark{3}}

\altaffiltext{1}{Key Laboratory of Solar Activity, National Astronomical Observatories, \\
Chinese Academy of Sciences, Beijing 100012, China; [yangliheng;zjun;liting]@bao.ac.cn}

\altaffiltext{2}{W. W. Hansen Experimental Physics Laboratory, Stanford University, Stanford, CA 94305, USA}

\altaffiltext{3}{Kwasan and Hida Observatories, Kyoto University, Kyoto 607-8471, Japan}

\begin{abstract}

We present detailed analysis of an extreme ultraviolet (EUV) wave and its interaction with active region (AR)
loops observed by the $\emph{Solar Dynamics Observatory}$/Atmospheric Imaging Assembly and the $\emph{Hinode}$
EUV Imaging Spectrometer (EIS). This wave was initiated from AR 11261 on 2011 August 4 and propagated at
velocities of 430--910 km\,s$^{-1}$. It was observed to traverse another AR and cross over
a filament channel on its path. The EUV wave perturbed neighboring AR loops and excited a disturbance that
propagated toward the footpoints of these loops. EIS observations of AR loops revealed that at the time
of the wave transit, the original redshift increased by about 3 km\,s$^{-1}$, while the original blueshift
decreased slightly. After the wave transit, these changes were reversed. When the EUV
wave arrived at the boundary of a polar coronal hole, two reflected waves were successively produced and part
of them propagated above the solar limb. The first reflected wave above the solar limb encountered a large-scale
loop system on its path, and a secondary wave rapidly emerged 144 Mm ahead of it at a higher speed. These
findings can be explained in the framework of a fast-mode magnetosonic wave interpretation for EUV waves,
in which observed EUV waves are generated by expanding coronal mass ejections.

\end{abstract}


\section{Introduction}

Large-scale wave-like coronal disturbances were first discovered by the Extreme Ultraviolet (EUV) Imaging
Telescope \citep[EIT;][]{del95} on board the $\emph{Solar and Helio-}$ $\emph{spheric Observatory}$
($\emph{SOHO}$), and were dubbed EIT or EUV waves \citep{mos97,tho98}. Generally, EUV waves
appear as broad, diffuse bright features followed by expanding dimming regions. Early $\emph{SOHO}$/EIT
observations indicated typical EUV wave velocities of 200--400 km\,s$^{-1}$ \citep{tho99,kla00,tho09}, while a
recent statistical study \citep{nit13} revealed a much higher average velocity of $\sim$600 km\,s$^{-1}$ for
more than 140 EUV waves observed at high cadence by the Atmospheric Imaging
Assembly \citep[AIA;][]{lem11} on board the $\emph{Solar Dynamics Observatory}$ \citep[$\emph{SDO}$;][]{pes12}.
EUV waves usually originate from flaring active regions (ARs) but have a strong association with coronal
mass ejections (CMEs) \citep[e.g.,][]{bie02,pat09,che09,ma11,che12}. Since their discovery, the nature of EUV waves
has been strongly debated. Several models have been proposed that can be grouped into wave, non-wave and hybrid
wave interpretations. The wave interpretation considers that EUV waves are real waves, including fast-mode
magnetohydrodynamic (MHD) waves or shock waves \citep[e.g.,][]{wan00,wu01,sch10}, slow-mode MHD waves \citep{wan09}
and soliton waves \citep{wil07}. Among them, the fast-mode MHD wave model is the most popular and supported by
many observations \citep{war01,ver08,kie09,gop09,patv09,pat09,liu11,zhe11,asa12}. The non-wave interpretation refers
to EUV waves as signatures of a current shell or successive restructuring of field lines during a CME \citep{del00,del08,che02,att07,sch11}.
The hybrid wave model points out that wave and non-wave phenomena, which might represent different
physical processes, could co-exist in a single event, and there is no need to develop a unified wave model to explain all the
observational characteristics of EUV waves \citep{zhu04,coh09,dow11,che02,che05}. More recent high
resolution observations are inclined to support this interpretation \citep{liu10,liu12,li12}. Comprehensive reviews of these models
and observations can be found in \citet{wil09}, \citet{war10}, \citet{gal11}, \citet{zhu11} and \citet{pat12}.

EUV waves have been observed to interact with various coronal structures in their paths, such as
coronal holes (CHs), ARs and coronal loops. Early observations revealed that EUV
waves could not travel across ARs \citep{wil99}, and stopped or partially intruded into
CHs \citep{tho98,tho99,ver06}. This was independently confirmed by numerical simulations of
\citet{wan00}, \citet{wu01} and \citet{ofm02}. EUV waves were observed by the EUV Imagers \citep[EUVI;][]{wue04}
on board the $\emph{Solar-Terrestrial Relations Observatory}$ \citep[$\emph{STEREO}$;][]{kai08} to be
reflected from CHs \citep{gop09}. This was further confirmed from higher cadence and sensitivity observations
by $\emph{SDO}$/AIA \citep{li12,she12}. Combining observations from $\emph{STEREO}$/EUVI and $\emph{SDO}$/AIA,
\cite{olm12} investigated an EUV wave and its interaction with a CH over the entire solar surface. They found that
part of the EUV wave transmitted through the CH, and the loop arcade at the CH boundary triggered a secondary wave, which
appeared to have been reflected. Secondary waves produced by distorted AR magnetic field have been simulated
by \citet{ofm02}, and then observationally confirmed by \citet{li12}. Additionally, EUV waves were reported to interact with coronal
loops \citep{tho98,del99,del00}. \citet{wil99} found in $\emph{Transition Region and Coronal Explorer}$ ($\emph{TRACE}$)
observations that an EUV wave propagated through diffuse, overarching coronal loops and triggered their transverse oscillations with a maximum
displacement of 6 Mm and velocity amplitudes of 15--20 km\,s$^{-1}$. Similar phenomena were observed by $\emph{SDO}$/AIA \citep{asc11,sch11,liu12,she12}.
In addition, filament oscillations can be triggered by the passage of an EUV wave \citep{oka04,her11} or Moreton wave \citep{gil08}.

Spectroscopic observations can provide plasma diagnostics of EUV waves and help to clarify their nature.
However, such observations are very rare due to the difficulty of aiming a narrow slit at the
location where a coronal wave will propagate \citep{zhu11}. \citet{har03} performed the first spectroscopic
analysis of an EUV wave event with the Coronal Diagnostic Spectrometer (CDS) instrument on board $\emph{SOHO}$.
In their observations, a weak wave front passed the CDS field of view but showed no significant
Doppler velocities ($\lesssim$ 10 km\,s$^{-1}$). By using the Extreme-Ultraviolet Imaging Spectrometer \citep[EIS;][]{cul07}
on board $\emph{Hinode}$, \citet{chen11} studied the interaction between an EIT wave and a coronal upflow region by
using $\emph{Hinode}$/EIS, and the upflow and non-thermal velocities were found to diminish after the passage of the wave front. They
suggested that this phenomenon implied changes of magnetic field orientation, and was consistent with the field line
stretching model of EUV waves. Recently, a unique data set was obtained during the $\emph{Hinode}$ Observing Plan HOP-180 by
placing a slit on the path of an EUV wave. Using these data, \citet{har11} found two redshift signatures corresponding to the wave pulses.
They propagated along the EIS slit with an average velocity of $\sim$500 km\,s$^{-1}$, which was similar to the velocity of the associated
wave front observed by $\emph{SDO}$/AIA. They considered that these redshifts might be signatures of plasma pushed downward and compressed by a coronal MHD
wave. In the followup work of the same event, \citet{ver11} found that a redshift of 20 km\,s$^{-1}$ was followed by a blueshift of $-$5 km\,s$^{-1}$,
indicating relaxation of the plasma behind the wave front. Both studies concluded that the observed wave was a coronal fast-mode MHD wave being
generated by the outgoing CME.

In this paper, we present detailed analysis of an EUV wave observed simultaneously by both $\emph{SDO}$/AIA and $\emph{Hinode}$/EIS that
sheds new light on its interaction with AR loops and CHs. We describe the observations in Section 2 and present analysis results in Sections 3 and 4,
followed by conclusions and discussion in Section 5.

\section{Observations and Data Analysis}

On 2011 August 4, AR 11261 produced an M9.3 class flare at the location of N16$^\circ$W49$^\circ$, which
started at 03:41~UT and peaked at 03:57~UT. Following the onset of the flare, we observed an EUV wave, a filament
eruption and a fast halo CME of 1315 km\,s$^{-1}$. The EUV wave was well observed by \emph{SDO}/AIA. AIA can
provide high resolution (1.5$''$), high cadence (12~s) full-disk images of the corona and transition region up to
0.5 R$_{\odot}$ above the solar limb. AIA images are taken in seven EUV passbands and three continuum bands, the former of
which are centered at specific lines: Fe {\sc xviii} (94~{\AA}), Fe {\sc viii},{\sc xxi} (131~{\AA}), Fe {\sc ix} (171~{\AA}),
Fe {\sc xii},{\sc xxiv} (193~{\AA}), Fe {\sc xiv} (211~{\AA}), He {\sc ii} (304~{\AA}), and Fe {\sc xvi} (335~{\AA}), covering
a wide temperature range from 6$\times$10$^{4}$ to 6$\times$10$^{7}$ K. This EUV wave was recorded by all of the seven
EUV channels. However, only four channels ( (171, 193, 211, and 335~{\AA}) were analyzed in detail, in which the EUV wave was more evident.

At about 04:01~UT, the eastern part of the EUV wave propagated to AR 11263 (N17$^\circ$W18$^\circ$) and
was captured by $\emph{Hinode}$/EIS. EIS observes the solar corona and upper transition region in two
EUV wavebands: 170--210~{\AA} and 250--290~{\AA}. It has a spatial resolution along the slit of 1$''$ pixel$^{-1}$ and
a spectral resolution of 0.0223~{\AA} pixel$^{-1}$, which permits Doppler velocity measurements better than 5 km\,s$^{-1}$.
Two spectral slits (1$''$ and 2$''$) provide high-resolution spectra and two imaging slots (40$''$ and 266$''$) provide
monochromatic images. In the present work, the 2$''$ slit was used to scan an area of 240$''$$\times$16$''$ with 25~s
exposure time, giving an averaged duration of about 3.5 min. We used the EIS data during the period
of 02:07--04:46~UT, including 19 continuous observation sequences from 03:45~UT to 04:46~UT, which allows us to study the plasma
behavior before, during, and after the EUV wave. However, the 04:39~UT and 04:40~UT sequences were only partially scanned and not
included in this study. Each sequence contained nine spectral windows, but our study mainly concentrated on two emission
lines (Fe {\sc xii} 195.12~{\AA} and Fe {\sc xiii} 202.04~{\AA}), whose maximum response temperatures were respectively 1.2 and 1.6~MK.
In addition, Si {\sc x} 258.37 and 261.04~{\AA} lines, which have the same maximum response temperature \textbf{as} the Fe {\sc xii} line,
were also chosen to make the density diagnosis.

We processed the raw EIS data by using the standard routine eis$\_$prep.pro in the SolarSoftWare
(SSW) packages \citep{fre98}, which corrected the data for cosmic ray hits, hot pixels, detector bias and dark
current. Then we used eis$\_$auto$\_$fit.pro with a single Gaussian model to obtain spectral intensities, line widths
and Doppler velocities. From the Si {\sc x} $\lambda$258.37/$\lambda$261.04 line pair, we derived the coronal
electron densities \citep{der97}.

AIA images were first differentially de-rotated to a common time at 05:00 UT and then running differenced by subtracting
the previous image in time. We adopted a semi-automatic approach \citep{pod05, liu10, li12} to track the wave propagation.
As shown in Figure 1, for the primary EUV wave on the solar disk, we defined the eruption center (N14$^\circ$W38$^\circ$)
as the new ``north pole'' from which three heliographic ``longitudinal'' sectors (labeled ``A''--``C'') of 10$^\circ$ wide were
selected. For each sector in an image, we averaged pixels in the perpendicular (``latitudinal'') direction and obtained
a one-dimensional profile as a function of spherical distance measured along the ``longitudinal'' great circle (thus accounting
for the Sun's sphericity). The distance step size along the sector was selected to correspond to AIA's pixel size of 0.6$''$.
Repeating this for a sequence of images and aligning the resulting profiles over time gave a two-dimensional time-distance plot.
Likewise, for the reflected EUV wave from AR 11263 (N11$^\circ$W9$^\circ$), we defined Sectors ``D''--``F'' of the same size centered there.
For the secondary wave above the southeast limb, we selected a cut of a right triangular shape (labeled ``G''), from which pixels were
averaged in the direction along the shortest side (from ``v1'' to ``v2'' in Figure 1).

The wave front was identified as a bright or dark track in the time-distance plot. To measure its velocity (acceleration), we first
determined that whether its slope was uniform. If it was uniform, we applied a linear (parabolic) fit to it. Otherwise, we divided it
according to its slope, and then applied a linear (parabolic) fit to the linear (non-linear) segment. For a certain wave front, dozens of
data points were chosen along the front of the wave pulse at a set time interval (marked by blue dotted lines in Figures 3--5). We performed
linear (parabolic) fit to these data points, repeated measurements ten times and then took average as the final velocity (acceleration).
The error was the standard deviation from the multiple measurements.

\section{AIA Observations of the EUV Wave}

\subsection{Overview of the EUV Wave}

Figure 2 displays the evolution of the EUV wave in the 211~{\AA} channel. The pre-event intensity map at 03:40:00~UT
clearly shows that there are a series of coronal loops (marked by ``L1'' in Figure 2(a))
connecting ARs 11261 and 11263. To the southeast of AR 11263, three small-scale coronal structures, where
reflected waves are observed, are identified with ``S1'', ``S2'' and ``S3''. The corresponding reflected waves
are marked by ``R1'' , ``R2'' and ``R3'' (see Figures 2(e) and (f)). The position of the EIS slit is superimposed on
this image (see the white box in Figure 2(a)), and is close to L1. The upper part of the slit was dominated
by the strong background emission in the core of AR 11263, so it is difficult to determine whether the wave passes
through it or not. Hence, only observations from the lower part of the slit, shown as the black box in Figures 2(b)--(f),
were used for our analysis. The running difference images in Figure 2 clearly show the propagation of the EUV wave.

As a typical feature of EUV wave events, erupting loops (marked by ``L2'' in Figure 2(b)) behind the EUV wave
were observed \citep[e.g.,][]{liu10,li12}. They appeared as semicircular bright arcades straddling AR 11261
and began to expand to the northwest at 03:48:00~UT. Three minutes later, a diffuse EUV wave pulse appeared
ahead of it. Initially, it mainly spread to the northwest, consistent with the orientation of L2 (see Animations 1 and 2).
In other directions, the EUV wave was relatively weak. We note that the loops to the south of AR 11261 (marked by ``L3'' in Figure 2(c) and Figure 3(b))
were disturbed, and it became more extended and brighter. Seen from the time-distance plot of Sector ``A'', some of L3 returned to their original
position after the EUV wave passed over, indicating oscillations triggered by the impact of the EUV wave
\citep[e.g.,][]{pat09,asc11,liu12}. By 03:54:24~UT, the EUV wave developed to a semi-circular shape (Figure 2(c)),
with a deformation in its eastern part, an indicator of the interaction of the EUV wave with L1, which became increasingly
evident as the EUV wave evolved.

To quantify the EUV wave kinematics, we used three time-distance slices along Sectors ``A''--``C''. As seen from Figures 2 and 3,
the EUV wave appeared as bright tracks in the hotter 193, 211 and 335~{\AA} channels, while a dark track in the cooler 171~{\AA} channel, suggesting
plasma heating \citep{wil99,liu10,liu12,lon11,ma11,li12}. In the 211 channel, the average propagation velocities of this EUV wave
were from (448$\pm$9) to (900$\pm$10) km\,s$^{-1}$ in the three directions. We note that in the same direction, the EUV wave displayed similar velocities
in the four channels and the velocity differences were almost within the error range except for the velocity of Sector ``B'' in the 171~{\AA} channel, which
was lower than those in the other three channels. This might be due to the fuzzy wave front. For Sector ``A'', the EUV wave exhibited a deceleration
between 400 Mm and 500 Mm from the eruption center (Figures 3(b)--(e)). A parabolic fit to the EUV wave within this distance revealed a deceleration
of (1010$\pm$80)--(1060$\pm$70) m\,s$^{-2}$. In sector ``C'', the EUV wave velocity decreased to
(270$\pm$20)--(301$\pm$6) km\,s$^{-1}$ at about 440 Mm from the eruption center (marked by the white dash-dotted line in Figures 4(a)--(e)).

In order to validate the measured EUV wave kinematics, we compared velocities and accelerations measured from Sector ``A'' in the running difference stack
plots (Figures 3(b)--(e)) with those in the base difference stack plots (Figures 3(b1)--(e1)), which had the base time at 03:40~UT. For the base difference
stack plots, the velocities and accelerations of the EUV wave were measured along the center other than the front of the wave pulse. It is noted that the
velocities of the same feature measured from the two methods had little difference. In addition, the deceleration between 400 Mm and 500 Mm from the eruption
center measured from the base difference stack plots was (980$\pm$70)--(1060$\pm$90) m\,s$^{-2}$, which was consistent with that measured from running
difference stack plots.

\subsection{Interaction of the EUV Wave with Nearby Inter-AR Loops}

In order to clearly describe the interaction between the EUV wave and L1, we divide the interaction into
two phases. The first phase was from 03:52:00 to 04:00:24~UT, that is, before the EUV wave reached
the apex of L1. During this phase, the EUV wave moved through L1 at a velocity of (430$\pm$10)--(448$\pm$9) km\,s$^{-1}$
(Figures 4(a)--(e); Animations 3 and 4), and the wave front had a deformation. Similar observations were reported
by \citet{wil99}, \citet{del99} and \citet{del00}.

The second phase ranged from 04:00:24 to 04:10:00~UT. At 04:00:24~UT, the EUV wave arrived at the apex of L1,
and excited a disturbance in L1 (see Figure 2(d)). Then, the EUV wave and the disturbance formed a clear bifurcation in the
space-time plot (Figure 2(d); the top and middle rows of Figure 4). Such bifurcation was obvious
in the 211, 193 and 335~{\AA} channels, but not observed in the 171~{\AA} channel. Simultaneously, the EUV wave pushed L1 forward (see Animation 3). L1
showed motions parallel to their magnetic field structure, different from the observations of \citet{wil99},
in which the disrupted coronal loops showed motions perpendicular to their magnetic field structure. Seen from the
time-distance plot of Sector ``C'', the velocity of L1 was about (64$\pm$3)--(97$\pm$7) km\,s$^{-1}$. This movement lasted about
10 min, leading to a maximum displacement of about 37--62 Mm. About 14 min later, another bright signal appeared at the apex of L1
(Figure 2(f) and Figures 4(a), (b), (d), and (e); Animation 3). It had a lower velocity of about (37$\pm$2)--(58$\pm$5) km\,s$^{-1}$, but
lasted a much longer time (about 15 min), giving a maximum displacement of about 32--57 Mm.

At the same time, the disturbance gradually propagated toward the footpoints of L1 (Animations 3 and 4). As seen from the
time-distance plot of Sector ``D'', it moved at a velocity of about (179$\pm$9)--(220$\pm$10) km\,s$^{-1}$. At 04:10:00~UT, it reached
the footpoints of L1.

\subsection{Passage of the EUV Wave Through AR 11264 and a Filament Channel}

At 03:55:36~UT, the EUV wave encountered a coronal bright point (see Figure 1, also indicated by the black dashed line in Figure 3).
Part of it just passed over this bright point while the rest of the wave front was deformed. Seen from
the time-distance plot of Sector ``A'', the EUV wave split into two branches, with a time delay of about 2 min.
We note that the two branches had similar velocities in the four AIA channels, (590$\pm$10)--(621$\pm$8) km\,s$^{-1}$
for the leading front, (590$\pm$20)--(600$\pm$20) km\,s$^{-1}$ for the trailing front. At 04:03~UT, the EUV wave passed through
AR 11264 (see Figure 1 and white dot-dashed line in Figure 3) rather than stopping at its boundary. As seen from the time-distance
plot of Sector ``A'', the velocity of the EUV wave in this AR was about (760$\pm$20)--(760$\pm$30) km\,s$^{-1}$, larger than
the original velocity, consistent with the result of \citet{li12}. This phenomenon can be clearly seen in the 211
and 193~{\AA} channels, weak in the 335~{\AA} channel, and not observed in the 171~{\AA} channel.

At 04:54:00~UT, the EUV wave arrived at the boundary of a filament channel, which was located between 330 Mm and 420 Mm from the eruption
center in Sector ``B'' (see Figure 1 and white dashed lines in Figure 3). It is noted that the EUV wave directly crossed
over the filament channel without any velocity variation.

\subsection{Reflected Waves from Coronal Structures}

At 04:08:48 UT, the southeast part of the EUV wave encountered some coronal structures in the quiet Sun,
S1 and S2 (Figure 2(a)), and produced reflected waves R1 and R2 that propagate toward the north (see Figure 2(e),
Figures 5(a) and (e), and Animation 2). Another reflected wave R3 was produced at S3 at 04:16:00 UT. As seen from the
time-distance plot of Sector ``C'' (Figures 4(a)--(e)), R3 appeared as brightening at 211, 193 and 335~{\AA}, but darkening at 171~{\AA}.
Its initial velocity was (300$\pm$10)--(330$\pm$10) km\,s$^{-1}$, slightly greater than that of the incident wave
(about (270$\pm$20)--(301$\pm$6) km\,s$^{-1}$). We note that R3 finally propagated into AR 11263, and thus we selected
Sectors ``E'' and ``F'' to track its late-stage propagation. The resulting time-distance plots are shown in Figure 5, giving R3's final
velocities ranging from (99$\pm$9) to (137$\pm$4) km\,s$^{-1}$.

Some intensity oscillations, as marked by the short blue line in Figure 1, occurred at S3 in the quiet Sun to the south
of AR 11263. They were captured both in Sector ``C'' at a distance of 500 Mm from the eruption center and in Sector ``E'' at 300 Mm,
as shown in the top panels of Figures 4 and 5. The oscillations, with a period of about 12 minutes, started at about the same
time as the reflected wave R3. In addition, as can be seen in Figure 5 (bottom), subsequent wave pulses appeared after the
initial R3 pulse, each delayed by about 12 minutes. This suggests that the multiple R3 wave pulses were generated by the
oscillations of the local loops at S3.

\subsection{Secondary Wave Excited by the Reflected Wave from a Polar CH}

At 04:22:00~UT, the EUV wave arrived at the boundary of the southern polar CH, and then a reflected wave was observed to
propagate to the northeast (see Animations 2 and 5). About 1 min later, part of this reflected wave appeared
above the solar limb (marked by ``RW1'' in Figures 6(b)--(e)). Seen from the stack plot of Sector ``G'', we note that it had a velocity
of 400$\pm$10 km\,s$^{-1}$, which was lower than that of the primary wave (621$\pm$8 km\,s$^{-1}$, see Figure 3(b)). At 04:30:00~UT, the reflected wave
encountered a large-scale loop system (marked by ``Loops'' in Figure 6(a)), and a secondary wave (marked by  ``SW'' in Figures 6(c), (d) and (e))
rapidly emerged 144 Mm ahead of it at a higher velocity of 510$\pm$20 km\,s$^{-1}$, in line with that reported by \citet{li12}. However,
instead of disappearing after the secondary wave was generated, the reflected wave continuously propagated forward. These are consistent with
secondary waves produced by distorted AR magnetic fields, which were simulated by \citet{ofm02}. At 04:30:24~UT, another reflected wave (marked
by ``RW2'' in Figures 6(c)--(e)) appeared above the solar limb at a velocity of 400$\pm$10 km\,s$^{-1}$. However, this reflected wave disappeared
after a certain distance before any interaction with the large-scale loop system was detected.

\section{Spectroscopic Analysis of the EUV Wave}

Figure 7 shows time sequences of the Doppler velocity and line width images for the Fe {\sc xiii} and Fe {\sc xii} lines.
At 04:00 UT when the EUV wave arrived, some red-shifted features appeared or were enhanced, being more evident in the hotter
Fe {\sc xiii} line. Simultaneously, the widths of both lines increased. After the EUV wave transit, there were some additional variations
in the Doppler velocity. A significant intensity enhancement was expected when the EUV wave swept through the EIS, but was not detected (see Figure 8(e)).

To study the variations of the Doppler velocity and line width in detail, we selected areas ``A'' and ``B'' (see Figure 7(a))
that were originally blue-shifted and red-shifted, respectively, and both showed distinct variations during the EUV wave passage.
We calculated the averaged Doppler velocity and line width in both areas for the two spectral lines, which are shown as a function
of time in Figure 8. At the time of the EUV wave transit, the red-shifted feature in ``B'' for the Fe {\sc xiii} line
increased by about 3 km\,s$^{-1}$ and formed a sharp peak (marked by ``vp1'' in Figure 8(a)), while the blue-shifted feature in ``A''
decreased slightly. Similar variations were also noticed in the Fe {\sc xii} line. Simultaneously, the line width in both ``A''
and ``B'' for the two lines clearly increased (marked by ``wp1'' in Figures 8(b) and (d)). For ``B'', it increased by
about 10~m{\AA} for the Fe {\sc xiii} line and about 7~m{\AA} for the Fe {\sc xii} line. The line width increases were somewhat
smaller in ``A'', by 6~m{\AA} and 5~m{\AA}, respectively.

There are two spikes in Doppler velocity following the initial peak at the time of the EUV wave arrival.
They appeared at around 04:10~UT (marked by ``vp2'' in Figures 8(a) and (c)) and 04:32~UT (marked by
``vp3'' in Figures 8(a) and (c)), respectively. The first spike seemed to be caused by the disturbance
of the EUV wave, while the second spike might be associated with the second peak in line width (marked by ``wp2'' in
Figures 8(b) and (d)), which appeared at 04:21~UT and might not be the result of the EUV wave.

In order to validate the above result, we also study the variation of the Doppler velocity and line width
between 02:07 and 03:08~UT. It is noticed that the variation of the Doppler velocity had no clear trend during this period.
For ``A'', the mean Doppler velocity was $-$1.3 km\,s$^{-1}$ for the Fe {\sc xiii} line and $-$3.0 km\,s$^{-1}$ for the
Fe {\sc xii} line (marked by the horizontal dash-dotted lines in Figures 8(a) and (c)). For ``B'', the mean Doppler velocity
was 2.9 km\,s$^{-1}$ for the Fe {\sc xiii} line and 1.6 km\,s$^{-1}$ for the Fe {\sc xii} line. Although the variation of
the line width during this period seemed to have periodicity, the variations were within 4~m{\AA} for the Fe {\sc xiii} line
and within 2~m{\AA} for the Fe {\sc xii} line.

Plasma densities, which were derived from the EIS Si {\sc x} line pair, as a function of time were presented in Figure 8(f).
It is noted that the changes of the plasma density were not obvious when the EUV wave passed over. The plasma density decreased by
about 2.9$\times$10$^{8}$ cm$^{-3}$ in ``A'' and 1.1$\times$10$^{8}$ cm$^{-3}$ in ``B'', which were within the variation range of
the plasma density between 02:07 and 03:08~UT. These observational results were similar to those given by \cite{ver11}, in which small
density changes associated with an EUV wave. We note that after the EUV wave transit, the plasma density reached a peak of
1.8$\times$10$^{10}$ cm$^{-3}$ in ``B'' and of 9.1$\times$10$^{9}$ cm$^{-3}$ in ``A'' , marked by ``dp2'' and ``dp1'', respectively.
The peak of the plasma density in ``A'' seemed to be associated with the peak of the Fe {\sc xii} intensity (marked by ``ip'') but
had a 14 min delay compared with vp1. The peaks of the plasma density and the intensity might be the result of disturbances in AR loops caused by
the EUV wave. Phase difference between Doppler velocity and intensity might provide an evidence for the existence of a slow-mode wave \citep{wang09},
indicating that the disturbances were probably slow-mode waves. Moreover, we note that dp1 and dp2 had a time difference of about 7 min,
which might imply that their oscillations are not in phase among different loops.

\section{Conclusions and Discussion}

We have presented detailed analysis of $\emph{SDO}$/AIA and $\emph{Hinode}$/EIS observations of an EUV wave associated with an M9.3
class flare and a fast halo CME, focusing on its interaction with local coronal loops. The main observational
results are summarized as follows.

1. The EUV wave had velocities from 430$\pm$10 to 910$\pm$10 km\,s$^{-1}$ in different directions. In the northeast
direction, it showed a clear deceleration of (1010$\pm$80)--(1060$\pm$70) m\,s$^{-2}$. Such deceleration
has been shown by multiple authors including \citet{war04}, \citet{lon11,long11}, \citet{muh11}, as well as \citet{ma11} and \citet{che12}.
According to the extensive statistical study of \citet{war11}, the initial wave speeds exceeding 320 km\,s$^{-1}$ showed pronounced deceleration.
They pointed out that the kinematic behavior of the decelerating wave was consistent with nonlinear large-amplitude waves or
shocks that propagated faster than the ambient fast-mode wave speed and subsequently slowed down due to decreasing amplitude.

2. The wave front first propagated through nearby inter-AR loops at a velocity of (430$\pm$10)--(448$\pm$9) km\,s$^{-1}$. When it
arrived at the apex of these loops, it excited a disturbance that finally propagated toward the footpoints
of these loops at a velocity of (179$\pm$9)--(220$\pm$10) km\,s$^{-1}$. Simultaneously, the EUV wave pushed these AR loops forward
with velocities of about (64$\pm$3)--(97$\pm$7) km\,s$^{-1}$ and a maximum displacement of about 37--62 Mm. About 14 min later, another
bright signal appeared at the apex of these loops. It had a lower velocity of about (37$\pm$2)--(58$\pm$5) km\,s$^{-1}$, but lasted a much longer time,
giving a maximum displacement of about 32--57 Mm. Such bright signal might be the result of the AR loop oscillation.
Previously, propagating disturbances in AR loops were reported by many researchers \citep[e.g.,][]{dem00,wan09,mci09,tia11}.
Such propagating disturbances were similar to our observed disturbance except that they were periodic. The nature of these propagating
disturbances \textbf{is} still an open question. Some researchers have suggested that the disturbances are mass flows \citep[e.g.,][]{mci09,tia11},
and the others considered them as slow-mode waves \citep[e.g.,][]{dem00,wang09}. More recently, \citet{ofm12} found that both slow-mode
waves and persistent upflows were present in the same impulsive events at the base of ARs based on the results of the 3D MHD model.

3. The EUV wave passed through AR 11264 and a filament channel on its path. This lends support to the fast-mode wave model of EUV waves,
but contradicts the prediction of the non-wave model \citep{che02} that an EUV wave should stop at a magnetic separatrix between
an AR or a filament channel and the rest of the corona. In addition, the velocity of the EUV wave increased by almost 200 km s$^{-1}$ within AR,
consistent with expected higher fast-mode wave speeds there. However, in contrast to the velocity increase of an EUV wave detected
within a flux rope cavity hosting a filament (Liu et al. 2012), there is no detectable velocity change at the filament channel
in our case. We speculate that the associated flux rope cavity in our case could have a similar fast-mode wave speed as the surrounding
corona or be situated at a lower height than the EUV wave in the corona.

4. When the EUV wave arrived at the boundary of a polar CH, two reflected waves were successively produced and part of them
propagated above the solar limb. The first reflected wave above the solar limb encountered a large-scale loop system on its path,
then a secondary wave rapidly emerged 144 Mm ahead of it at a higher speed. This is the first time that a secondary wave produced
by a reflected wave is observed. It is consistent with secondary waves produced by distorted AR magnetic fields, which were simulated
by \citet{ofm02}. Recently, \citet{li12} studied a global EUV wave on 2011 June 7. They found that when the EUV wave arrived at an AR on
its path, the primary EUV wave apparently disappeared and a secondary wave rapidly reemerged within 75 Mm of the AR boundary at a similar
speed. Similar to their observation, we also observed a secondary wave at a certain distance away from the initial wave. However, our
initial wave continuously propagated forward rather than disappearing. It might indicate that the observed secondary wave might be a newly
generated wave due to the interaction between the reflected wave and the large-scale loop system. In addition, we observed two reflected waves,
which might be triggered by the oscillations of the coronal hole boundary upon the impact of the primary EUV wave and the 8 min delay
of the two reflected waves may indicate the oscillation period.

5. EIS observations of AR loops in AR 11263 revealed that at the time of the wave transit, the original red-shifted
feature had an increase of about 3 km\,s$^{-1}$, and the original blue-shifted feature slightly weakened. After the wave
transit, these changes were reversed. Our results are similar to the recent studies using $\emph{Hinode}$/EIS and
$\emph{SDO}$/AIA observations \citep{har11,ver11}. It can be explained by the scenario of a fast-mode wave
model as described by \cite{uch68}. When the EUV wave encountered the AR loops, it would provide a downward pulse to these
loops \citep[e.g.,][]{liu12}. After the EUV wave transit, these disturbed loops restored to its original state. In addition,
we did not detect any evident intensity enhancement among the available EIS lines when the EUV wave swept through the EIS slit.
This is consistent with the findings of \citet{chen11} who concluded that the variation of line intensity during the wave
propagation was within the fitting error, so it is difficult to distinguish such changes from spectroscopic observations.
Coronal magnetic field strength of quiet Sun can be calculated through coronal seismology by measuring the wave speed,
coronal density and temperature \citep{wes11,lon13}. In our event, though the EIS slit was located in AR 11263, the area in the lower part
of the slit was relatively quiet. Hence, the local wave velocity and coronal density can be used to estimate the local magnetic field
strength. In order to determine the local wave velocity, we selected a 10$^\circ$ wide sector, which started from the eruption center and
intersected the lower part of the EIS slit, and derived a time-distance plot in the 193~{\AA} channel, who had the same maximum response
temperature (about 1.2 MK) with the Si {\sc x} line. The wave velocity was estimated to be 410$\pm$20 km\,s$^{-1}$. We note that when
the EUV wave swept through the EIS slit, the plasma density was 5.5$\times$10$^{8}$ cm$^{-3}$ in ``A'' and 4.4$\times$10$^{8}$ cm$^{-3}$
in ``B''. The local magnetic field strength can be calculated using B=(4$\pi$n(mv$_{fm}$$^{2}$-$\gamma$k$_{B}$T))$^{1/2}$, which was
deduced by \citet{lon13}. B is the magnetic field strength, v$_{fm}$ is the wave speed, n is plasma density, m is the proton mass, $\gamma$=5/3
is the adiabatic index, k$_{B}$ is the Boltzmann constant and T is the peak emission temperature of the density sensitive lines.
The calculated local magnetic field strength is 1.6$\pm$0.2 G in ``A'' and 1.3$\pm$0.1 G in ``B''.

In summary, we find a series of phenomena associated with the interaction of incident or reflected waves with coronal loops, such as a
disturbance in AR loops excited by the EUV wave and a secondary wave generated by the reflected wave from a polar CH. These observational
findings can be explained by the fast-mode magnetoacoustic interpretations for EUV waves, in which the observed waves are generated by expanding CMEs.

\acknowledgments

We thank the anonymous referee for the helpful suggestions and comments. We also thank the
$\emph{Hinode}$/EIS and $\emph{SDO}$/AIA teams for providing the wonderful data. $\emph{Hinode}$ is
a Japanese mission developed and launched by ISAS/JAXA, with NAOJ as domestic partner and NASA
and STFC (UK) as international partners. It is operated by these agencies in co-operation with ESA and NSC (Norway).
The work is supported by the National Natural Science Funds for Distinguished Young Scholar (11025315), the National
Basic Research Program of China under grant 2011CB811403, National Natural Science Foundations of
China (10921303, 11103075 and 11203037), and the CAS Project KJCX2-EW-T07.

\appendix

\clearpage

\begin{figure}
\epsscale{0.9}
\plotone{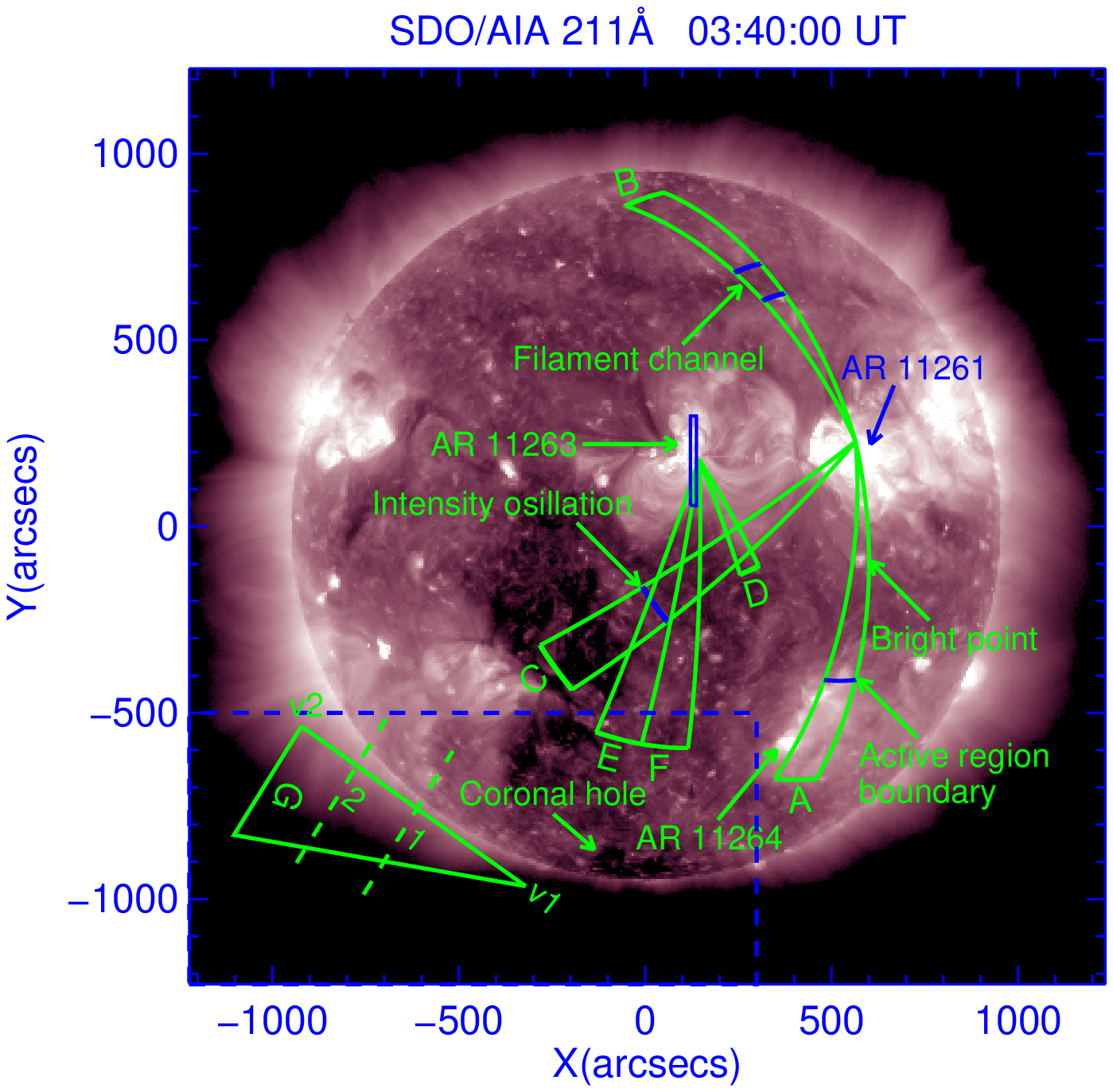}
\caption{Six 10$^\circ$ wide sectors (``A''--``F'') and one triangular sector (``G'') displayed on an
$\emph{SDO}$/AIA 211~{\AA} full-disk image which are used to obtain space-time plots in Figures 3--6. \label{fig1}}
\end{figure}

\clearpage

\begin{figure}
\epsscale{1.0}
\plotone{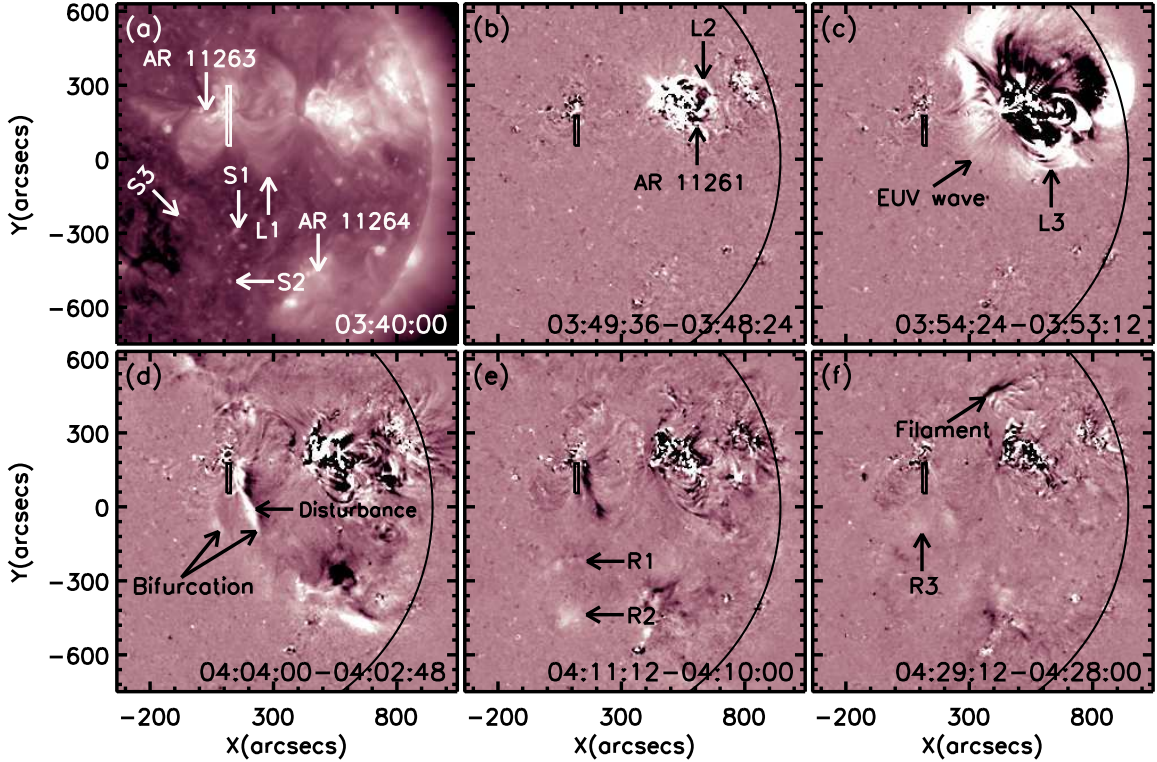}
\caption{Sequences of $\emph{SDO}$/AIA 211~{\AA} direct (a) and running difference images ((b)--(f)) showing
the propagation of the EUV wave. The white box in (a) indicates the position of the $\emph{Hinode}$/EIS
spectrometer slit. The field of view (FOV) of Figure 7 is marked by black boxes in (b)--(f). ``L1'' denotes
AR loops connecting AR 11261 and AR 11263. ``L2'' and ``L3'' are expanding loops during the EUV wave transit.
``S1'', ``S2'' and ``S3'' represent three coronal structures producing reflected waves, and ``R1'', ``R2'' and ``R3''
correspond to the reflected waves. The black curve in (b)--(f) marks the limb of the solar disk. The FOV
is 1380$''$$\times$1380$''$. \label{fig2}}
\end{figure}

\clearpage
\begin{figure}
\epsscale{0.9}
\plotone{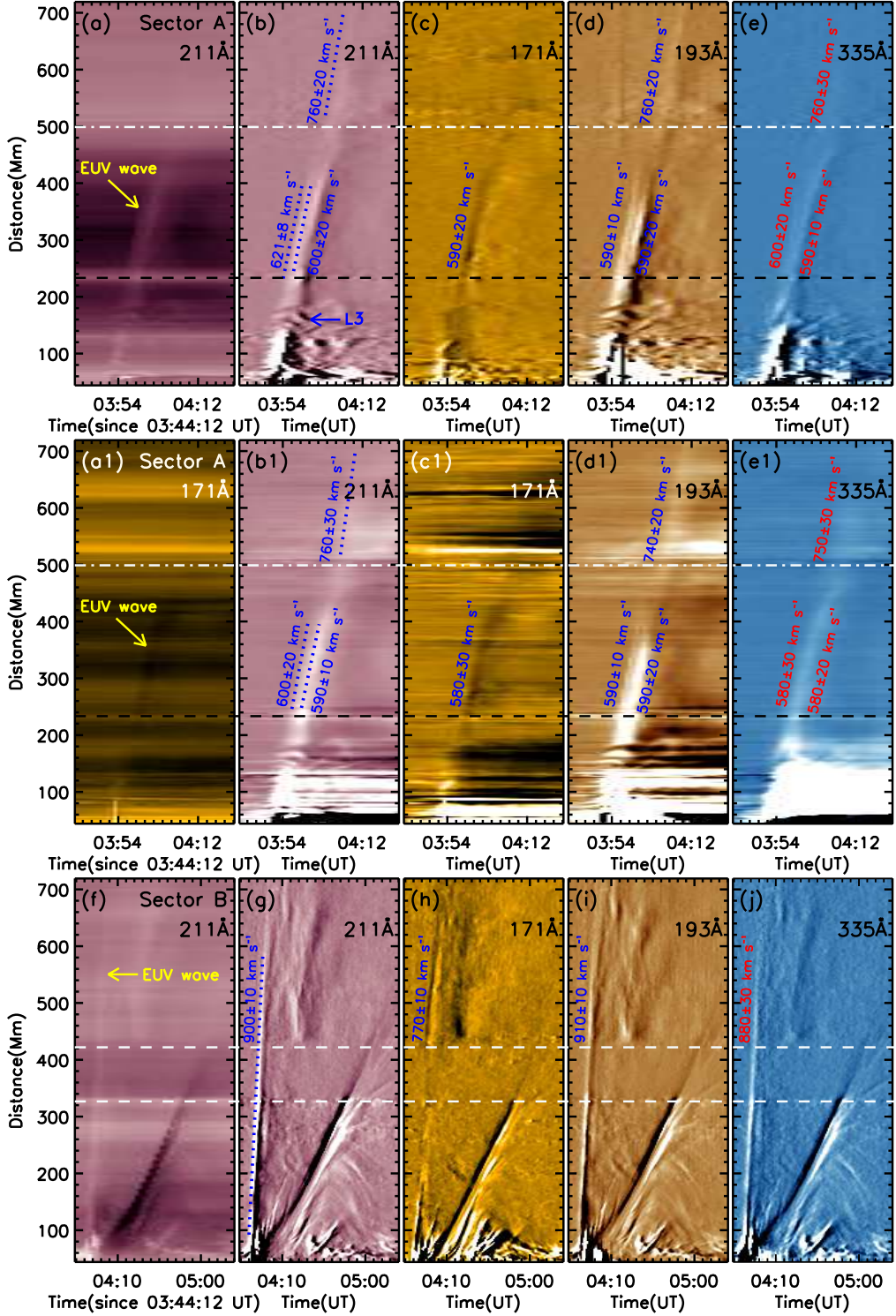}
\caption{Original ((a) and (f)), base ((b1)--(e1)) and running ((b)--(e) and (g)--(j)) difference space-time plots along
Sectors ``A'' and ``B'' at 211, 171, 193 and 335~{\AA}. Original space-time plot along Sector ``A'' at 171 ~{\AA} is shown in (a1).
``L3'' denotes expanding loops during the EUV wave transit. \label{fig3}}

\end{figure}

\clearpage
\begin{figure}
\epsscale{0.9}
\plotone{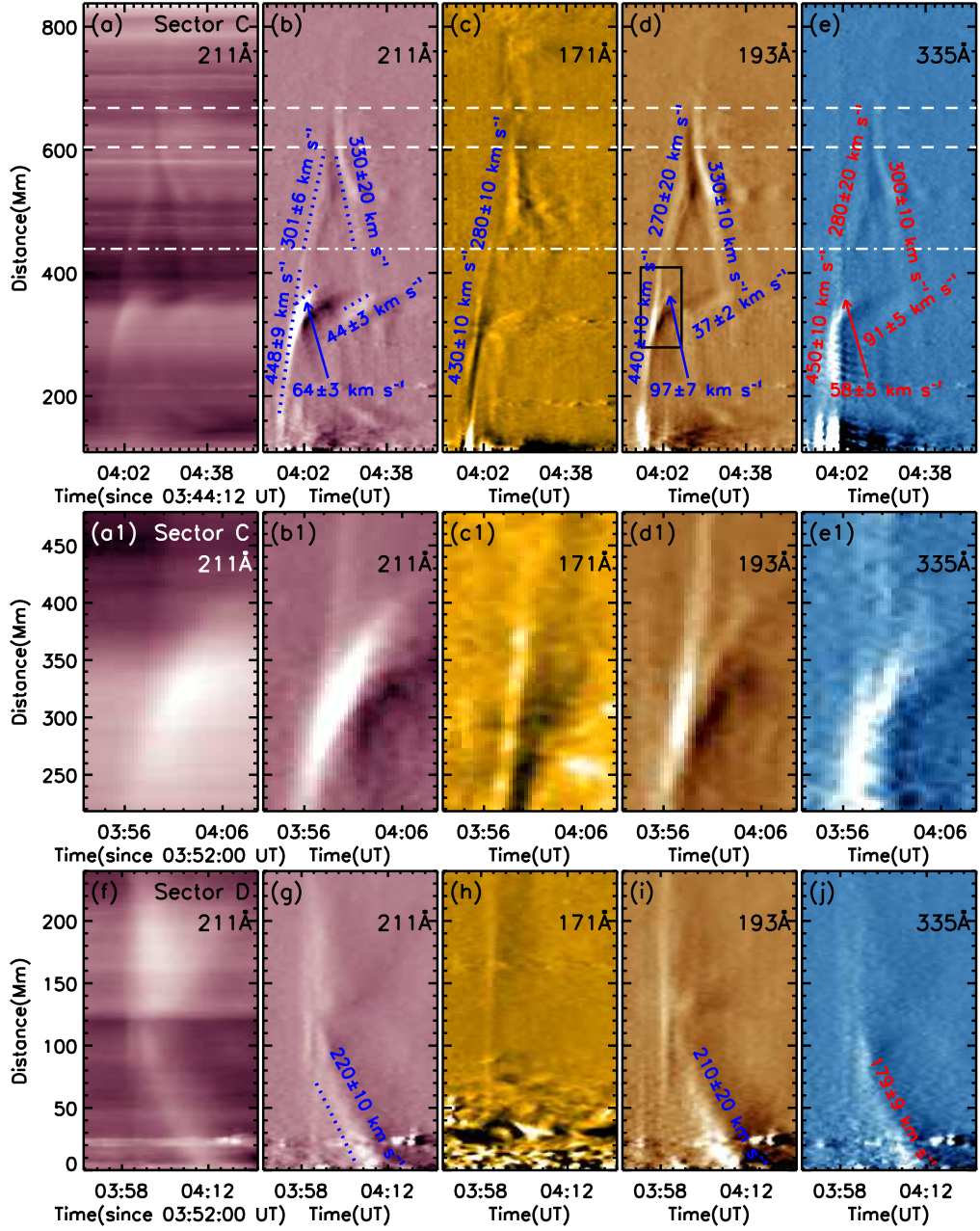}
\caption{Original (Column 1) and running difference (Columns 2--5) space-time plots along Sectors ``C'' and ``D'' at 211, 171, 193
and 335~{\AA}. The black box in panel (d) indicates the FOV of panels (a1)--(e1). \label{fig4}}
\end{figure}

\clearpage
\begin{figure}
\epsscale{0.9}
\plotone{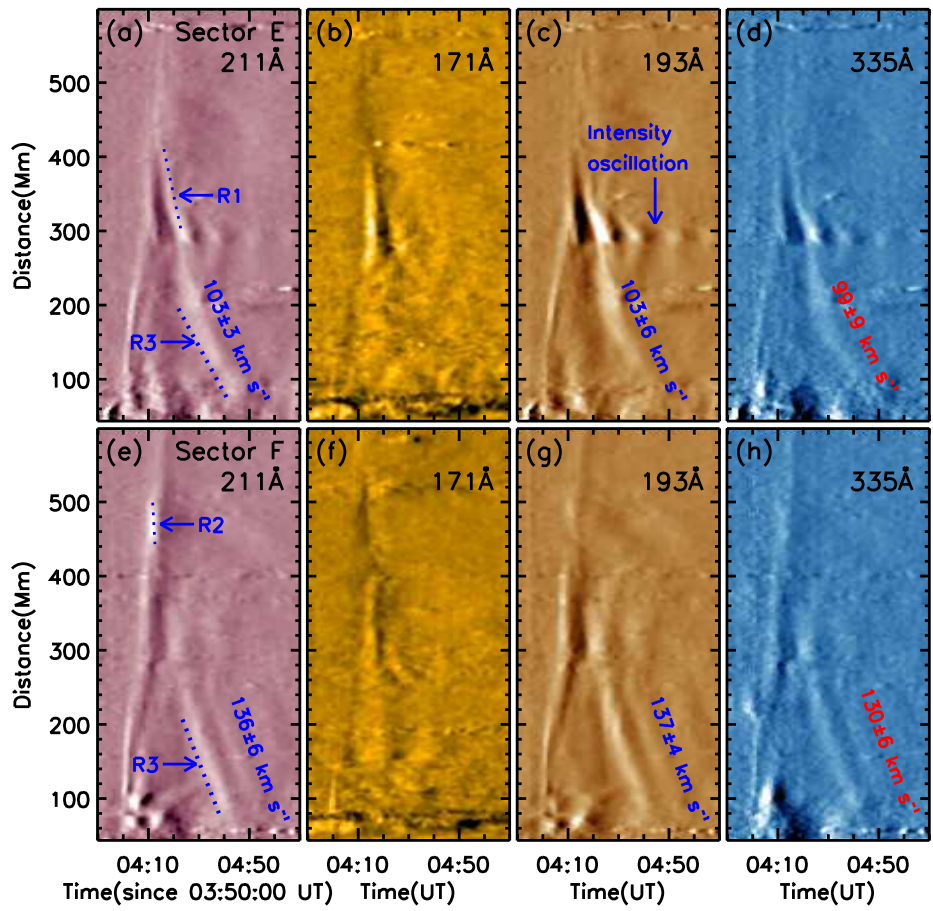}
\caption{Running difference space-time plots along Sectors ``E'' and ``F'' at 211, 171, 193 and 335~{\AA}. \label{fig5}}
\end{figure}

\clearpage
\begin{figure}
\epsscale{0.9}
\plotone{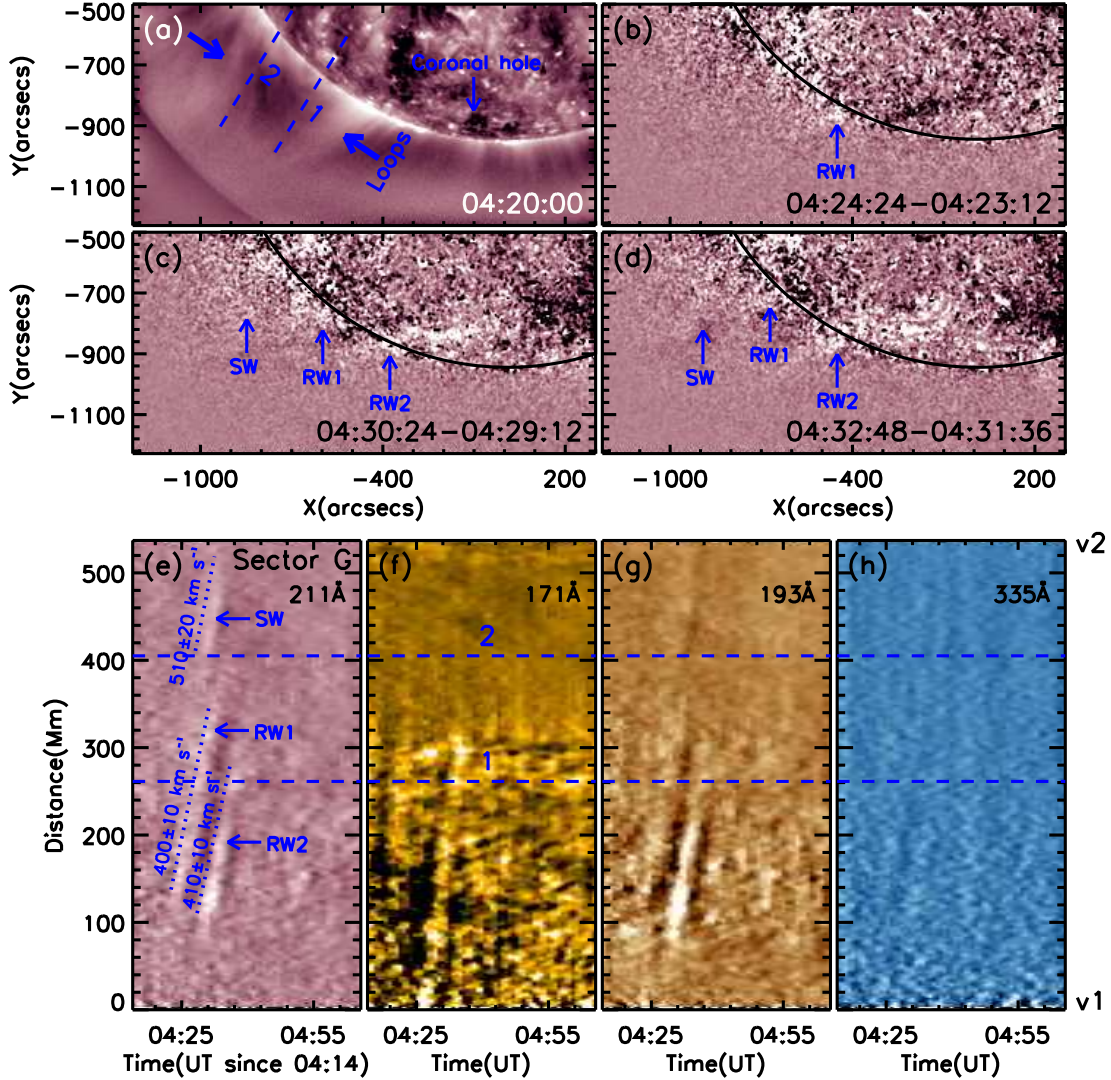}
\caption{Sequences of $\emph{SDO}$/AIA 211~{\AA} direct (a) and running difference ((b)--(d)) images showing the two reflected waves of
the southern polar CH and the newly produced secondary wave. The FOV is indicated by the rectangle in Figure 1.
Running difference space-time plots ((e)--(h)) are along the Sector ``G'' at 211, 171, 193 and 335~{\AA}. \label{fig6}}
\end{figure}

\clearpage
\begin{figure}
\epsscale{0.9}
\plotone{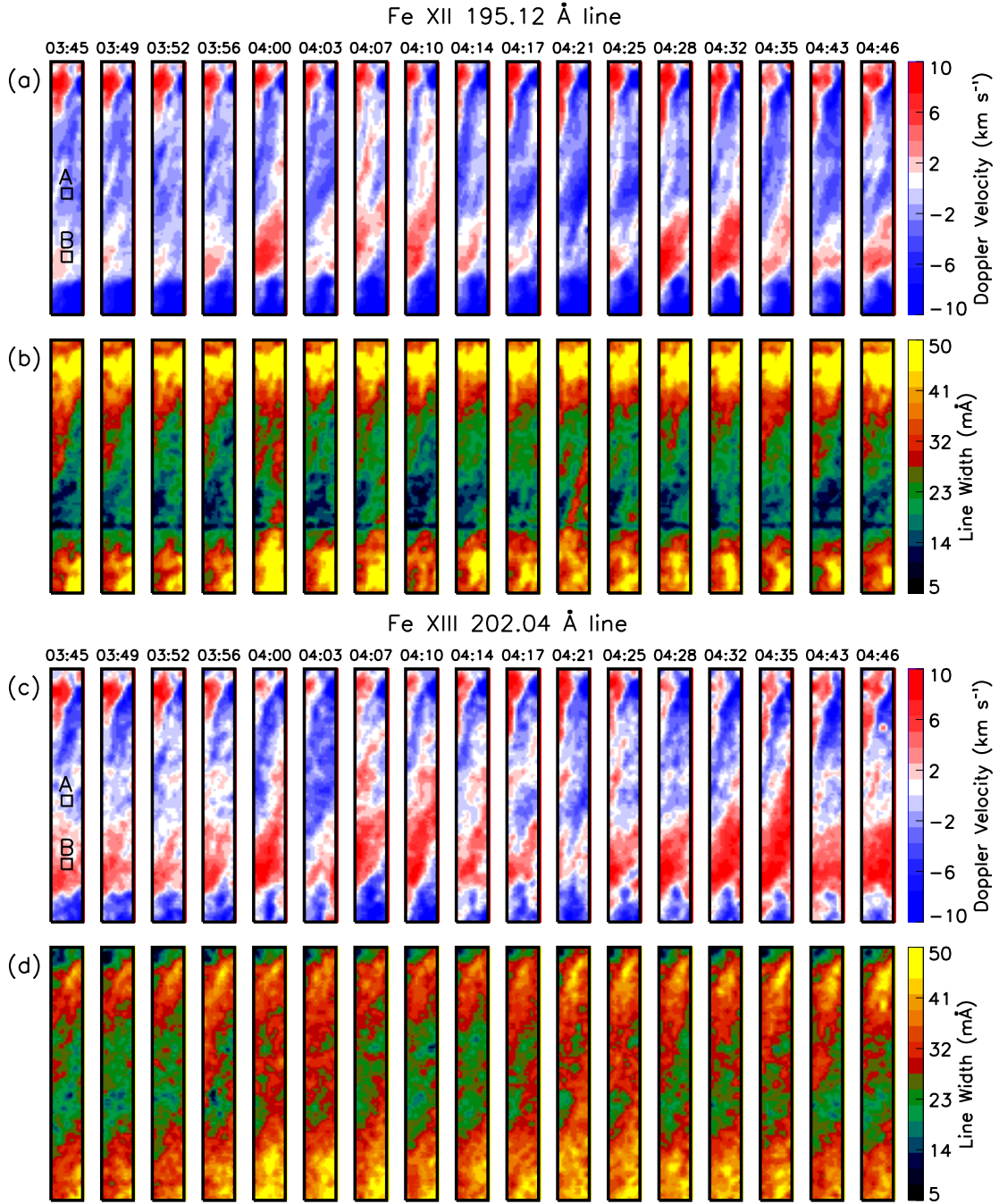}
\caption{Time sequences of Doppler velocity ((a) and (c)) and line width ((b) and (d)) images for the Fe
{\sc xii} 195.12~{\AA} and Fe {\sc xiii} 202.04~{\AA} lines. ``A'' and ``B'' are two regions where
averaged Doppler velocity and line width are calculated in Figure 8. The FOV is 16$''$$\times$120$''$. \label{fig7}}
\end{figure}

\clearpage
\begin{figure}
\epsscale{0.9}
\plotone{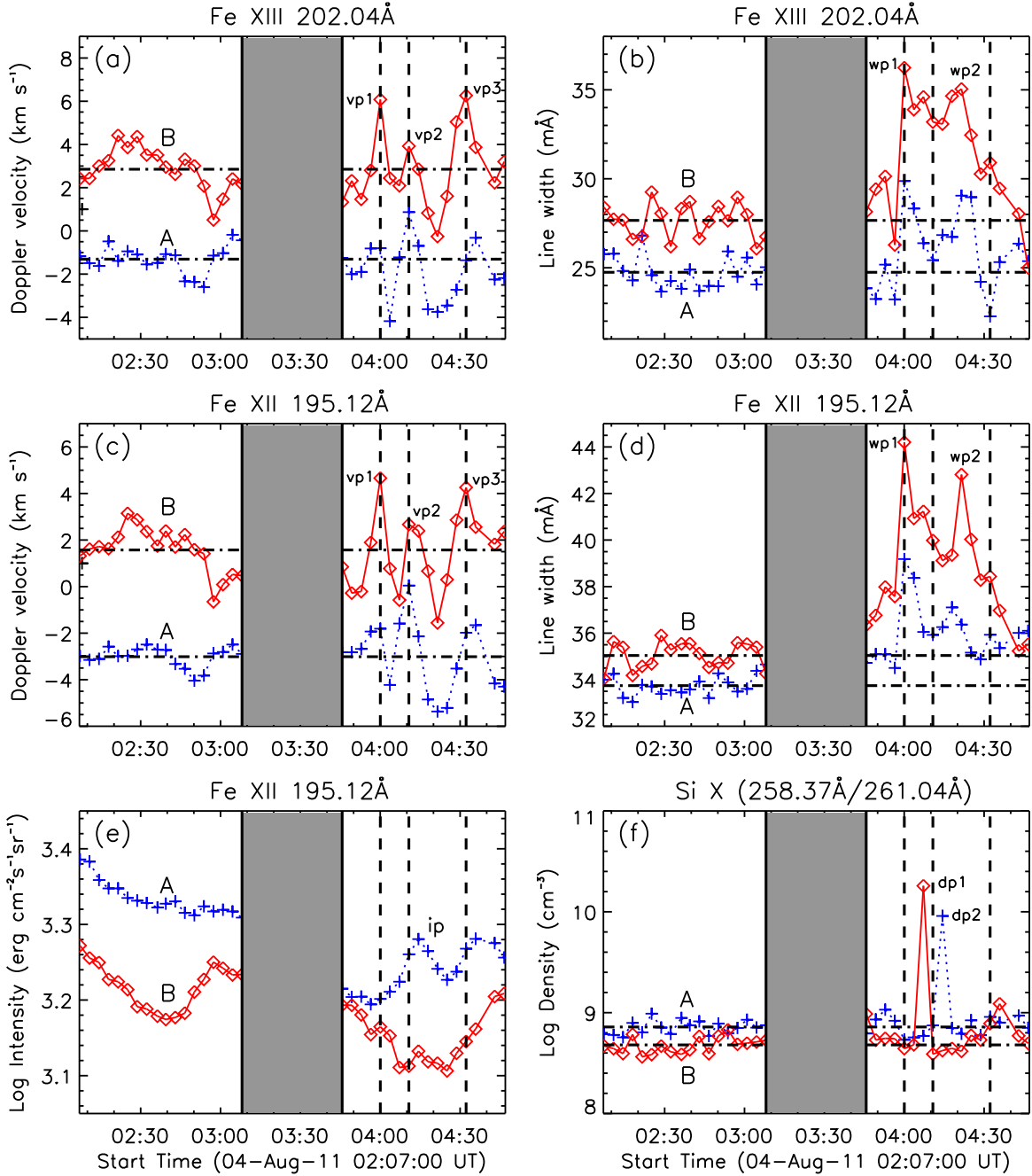}
\caption{Averaged Doppler velocity ((a) and (c)), line width ((b) and (d)), line intensity (e) and plasma density (f) (calculated within ``A''
and ``B'' in Figure 7) as a function of time. ``vp1'', ``vp2'' and ``vp3'' represent the three peaks of Doppler velocity.
``wp1'' and ``wp2'' denote the two peaks of line width. ``dp1'' and ``dp2'' represent the two peaks of density, and ``ip'' denotes
the peak of line intensity. The gray areas indicate that there are no data in the related time range. The horizontal dark dash-dotted lines represent
the mean values of the Doppler velocity, line width and plasma density during the period between 02:07 and 03:08~UT. \label{fig8}}
\end{figure}


\begin{thebibliography}{}

\bibitem[Asai et al.(2012)]{asa12} Asai, A., Ishii, T.~T., Isobe, H., et al. 2012, \apjl, 745, L18
\bibitem[Aschwanden \& Schrijver(2011)]{asc11} Aschwanden, W., \& Schrijver C.~J. 2011, \apj, 736, 102
\bibitem[Attill et al.(2007)]{att07} Attill, G.~D.~R., Harra, L.~K., van Driel-Gesztelyi, L.,
     \& D$\acute{e}$moulin, P. 2007, \apj, 656, L101
\bibitem[Biesecker et al.(2002)]{bie02} Biesecker, D.~A., Myers, D.~C., Thompson, B.~J., Hammer, D.~M., \& Vourlidas, A. 2002, \apj, 569, 1009
\bibitem[Chen et al.(2011)]{chen11} Chen, F., Ding, M.~D., Chen, P.~F., \& Harra, L.~K. 2011, \apj, 740, 116
\bibitem[Chen(2009)]{che09} Chen, P.~F. 2009, \apjl, 698, L112
\bibitem[Chen et al.(2005)]{che05} Chen, P.~F., Fang, C., \& Shibata, K. 2005, \apj, 622, 1202
\bibitem[Chen et al.(2002)]{che02} Chen, P.~F., Wu, S.~T., Shibata, K., \& Fang, C. 2002, \apjl, 572, L99
\bibitem[Cheng et al.(2012)]{che12} Cheng, X., Zhang, J., Olmedo, O., et al., 2012, \apjl, 745, L5
\bibitem[Cohen et al.(2009)]{coh09} Cohen, O., Attrill, G.~D.~R., Manchester, W.~B., IV, \& Wills-Davey, M.~J. 2009,
     \apj, 705, 587
\bibitem[Culhane et al.(2007)]{cul07} Culhane, J.~L., Harra, L.~K., James, A.~M., et al. 2007, \solphys, 243, 19
\bibitem[Delaboudini$\grave{e}$re et al.(1995)]{del95} Delaboudini$\grave{e}$re, J.-P., Artzner, G.~E.,
     Brunaud, J., et al. 1995, \solphys, 162, 291
\bibitem[Delann$\acute{e}$e(2000)]{del00} Delann$\acute{e}$e, C. 2000, \apj, 545, 512
\bibitem[Delann$\acute{e}$e \& Aulanier(1999)]{del99} Delann$\acute{e}$e, C., \& Aulanier, G. 1999, \solphys, 190, 107
\bibitem[Delann$\acute{e}$e et al.(2008)]{del08} Delann$\acute{e}$e, C., T$\ddot{o}$r$\ddot{o}$k, T., Aulanier, G.,
     \& Hochedez, J.-F. 2008, \solphys, 247, 123
\bibitem[De Moortel et al.(2000)]{dem00} De Moortel, I., Ireland, J., \& Walsh, R.~W. 2000, \aap, 355, L23
\bibitem[Dere et al.(1997)]{der97} Dere, K.~P., Landi, E., Mason, H.~E., Monsignori Fossi, B.~C., \& Young, P.~R. 1997, \aaps, 125, 149
\bibitem[Downs et al.(2011)]{dow11} Downs, C., Roussev, I.~I., van der Holst, B., et al. 2011, \apj, 728, 2
\bibitem[Freeland \& Handy(1998)]{fre98} Freeland, S.~L., \& Handy, B.~N. 1998, \solphys, 182, 497
\bibitem[Gallagher \& Long(2011)]{gal11} Gallagher, P.~T., \& Long, D.~M. 2011, \ssr, 158, 365
\bibitem[Gilbert et al.(2008)]{gil08} Gilbert, H.~R., Daou, A.~G., Young, D., Tripathi, D., \& Alexander, D. 2008, \apj, 685, 629
\bibitem[Gopalswamy et al.(2009)]{gop09} Gopalswamy, N., Yashiro, S., Temmer, M., et al. 2009, \apjl, 691, L123
\bibitem[Harra \& Sterling(2003)]{har03} Harra, L.~K., \& Sterling, A.~C. 2003, \apj, 587, 429
\bibitem[Harra et al.(2011)]{har11} Harra, L.~K., Sterling, A.~C., G$\ddot{o}$m$\ddot{o}$ry, P., \& Veronig A. 2011,
     \apjl, 737, L4
\bibitem[Hershaw et al.(2011)]{her11} Hershaw, J., Foullon, C., Nakariakov, V.~M., \& Verwichte, E. 2011, \aap, 531, A53
\bibitem[Kaiser et al.(2008)]{kai08} Kaiser, M.~L., Kucera, T.~A., Davila, J.~M., et al. 2008, \ssr, 136, 5
\bibitem[Kienreich et al.(2009)]{kie09} Kienreich, I.~W., Temmer, M., \& Veronig, A.~M. 2009, \apjl, 703, L118
\bibitem[Klassen et al.(2000)]{kla00} Klassen, A., Aurass, H., Mann, G., \& Thompson, B.~J. 2000, \aaps, 141, 357
\bibitem[Lemen et al.(2011)]{lem11} Lemen, J.~R., Title, A.~M., Akin, D.~J., et al. 2012, \solphys, 275, 17
\bibitem[Li et al.(2012)]{li12} Li, T., Zhang, J., Yang, S.~H., \& Liu, W. 2012, \apj, 746, 13
\bibitem[Liu et al.(2010)]{liu10} Liu, W., Nitta, N.~V., Schrijver, C.~J., Title, A.~M., \& Tarbell, T.~D. 2010, \apjl, 723, L53
\bibitem[Liu et al.(2012)]{liu12} Liu, W., Ofman, L., Nitta, N.~V., et al. 2012, \apj, 753, 52
\bibitem[Liu et al.(2011)]{liu11} Liu, Y., Luhman, J.~G., Bale, S.~D., \& Lin, R.~P. 2011, \apj, 734, 84
\bibitem[Long et al.(2011a)]{lon11} Long, D.~M., DeLuca, E.~E., \& Gallagher, P.~T. 2011a, \apjl, 741, L21
\bibitem[Long et al.(2011b)]{long11} Long, D.~M., Gallagher, P.~T., McAteer, R.~T.~J., \& Bloomfield, D.~S. 2011b, \aap, 531, 42
\bibitem[Long et al.(2013)]{lon13} Long, D.~M., Williams, D.~R., R$\acute{e}$gnier, S., \& Harra, L.~K. 2013, in press(arXiv:1305.5169)
\bibitem[Ma et al.(2011)]{ma11} Ma, S.~L., Raymond, J.~C., Golub, L., et al. 2011, \apj, 738, 160
\bibitem[McIntosh \& De Pontieu(2009)]{mci09} McIntosh, S.~W., \& De Pontieu, B. 2009, \apjl, 706, L80
\bibitem[Moses et al.(1997)]{mos97} Morses, D., Clette, F., Delaboudini$\grave{e}$re J.-P., et al. 1997, \solphys, 175, 571
\bibitem[Muhr et al.(2011)]{muh11} Muhr, N., Veronig, A.~M., Kienreich, I.~W., Temmer, M., \& Vr$\breve{s}$nak, B. 2011, \apj, 739, 89
\bibitem[Nitta et al.(2013)]{nit13} Nitta, N.~V., Schrijver, C.~J., Title, A.~M., \& Liu, W. 2013, \apj, submitted
\bibitem[Ofman \& Thompson(2002)]{ofm02} Ofman, L., \& Thompson, B.~J. 2002, \apj, 574, 440
\bibitem[Ofman et al.(2012)]{ofm12} Ofman, L., Wang, T.~J., \& Davila J.~M. 2012, \apj, 754, 111
\bibitem[Okamoto et al.(2004)]{oka04} Okamoto, T.~J., Nakai, H., Keiyama, A., et al. 2004, \apj, 608, 1124
\bibitem[Olmedo et al.(2012)]{olm12} Olmedo, O., Vourlidas, A., Zhang, J., \& Cheng X. 2012, \apj, 756, 143
\bibitem[Patsourakos \& Vourlidas(2009)]{patv09} Patsourakos, S., \& Vourlidas, A. 2009, \apjl, 700, L182
\bibitem[Patsourakos \& Vourlidas(2012)]{pat12} Patsourakos, S., \& Vourlidas, A. 2012, \solphys, 281, 187
\bibitem[Patsourakos et al.(2009)]{pat09} Patsourakos, S., Voulidas, A., Wang, Y.-M., Stenborg, G., \& Thernisien, A.
     2009, \solphys, 259, 49
\bibitem[Pesnell et al.(2012)]{pes12} Pesnell, W.~D., Thompson, B.~J., \& Chamberlin, P.~C. 2012, \solphys, 275, 3
\bibitem[Podladchikova \& Berghmans(2005)]{pod05} Podladchikova, O., \& Berghmans, D. 2005, \solphys, 228, 265
\bibitem[Schmidt \& Ofman(2010)]{sch10} Schmidt, J.~M., \& Ofman, L., 2010, \apj, 713, 1008
\bibitem[Schrijver et al.(2011)]{sch11} Schrijver, C.~J., Aulanier, G., Title, A.~M., Pariat, E., \& Delann$\acute{e}$e, C.
     2011, \apj, 738, 167
\bibitem[Shen \& Liu(2012)]{she12} Shen, Y.~D., \& Liu Y., 2012, \apj, 754, 7
\bibitem[Tian et al.(2011)]{tia11} Tian, H., McIntosh, S.~W., De Pontieu, B., et al. 2011, \apj, 738, 18
\bibitem[Thompson et al.(1999)]{tho99} Thompson, B.~J., Gurman, J.~B., Neupert, W.~M., et al. 1999, \apjl, 517, L151
\bibitem[Thompson et al.(1998)]{tho98} Thompson, B.~J., Plunkett, S.~P., Gurman, J.~B., et al. 1998, \grl, 25, 2465
\bibitem[Thompson \& Myers(2009)]{tho09} Thompson, B.~J., \& Myers, D.~C. 2009, \apjs, 183, 225
\bibitem[Veronig et al.(2011)]{ver11} Veronig, A.~M., G$\ddot{o}$m$\ddot{o}$ry, P., Kienreich, L.~W., et al. 2011, \apjl, 743, L10
\bibitem[Veronig et al.(2008)]{ver08} Veronig, A.~M., Temmer, M., \& Vr$\check{s}$nak, B. 2008, \apjl, 681, L113
\bibitem[Veronig et al.(2006)]{ver06} Veronig, A.~M., Temmer, M., Vr$\check{s}$nak, B., \& Thalmann, J.~K. 2006,
     \apj, 647, 1466
\bibitem[Uchida(1968)]{uch68} Uchida, Y. 1968, \solphys, 4, 30
\bibitem[Wang et al.(2009a)]{wan09} Wang, H., Shen, C., \& Lin, J. 2009a, \apj, 700, 1716
\bibitem[Wang et al.(2009b)]{wang09} Wang, T.~J., Ofman, L., \& Davila, J.~M. 2009b, \apj, 696, 1448
\bibitem[Wang(2000)]{wan00} Wang, Y.-M. 2000, \apjl, 543, L89
\bibitem[Warmuth(2010)]{war10} Warmuth, A. 2010, Advances in Space Research, 45, 527
\bibitem[Warmuth \& Mann(2011)]{war11} Warmuth, A., \& Mann, G. 2011, \aap, 532, A151
\bibitem[Warmuth et al.(2001)]{war01} Warmuth, A., Vr$\check{s}$nak, B., Aurass, H., \& Hanslmeier, A. 2001,
     \apjl, 560, L105
\bibitem[Warmuth et al.(2004)]{war04} Warmuth, A., Vr$\breve{s}$nak, B., Magdaleni$\acute{c}$, J., Hanslmeier, A., \& Otruba, W. 2004, \aap, 418, 1117
\bibitem[West et al.(2011)]{wes11} West, M.~J., Zhukov, A.~N., Dolla, L., \& Rodriguez, L. 2011, \apj, 730, 122
\bibitem[Wills-Davey \& Attrill(2009)]{wil09} Wills-Davey, M.~J., \& Attrill, G.~D.~R. 2009, \ssr, 149, 325
\bibitem[Wills-Davey et al.(2007)]{wil07} Wills-Davey, M.~J., DeForest, C.~E., \& Stenflo, J.~O. 2007, \apj, 664, 556
\bibitem[Wills-Davey \& Thompson(1999)]{wil99} Wills-Davey, M.~J., \& Thompson, B.~J. 1999, \solphys, 190, 467
\bibitem[Wu(2001)]{wu01} Wu, S.~T., Zheng, H., Wang, S., et al. 2001, \jgr, 106, 25089
\bibitem[Wuelser et al.(2004)]{wue04} Wuelser, J.~P., Lemen, J.~R., Tarbell, T.~D., et al. 2004, \procspie, 5171, 111
\bibitem[Zheng et al.(2011)]{zhe11} Zheng, R.~S., Jiang, Y.~C., Hong, J.~C., et al. 2011, \apjl, 739, L39
\bibitem[Zhukov(2011)]{zhu11} Zhukov, A.~N. 2011, Journal of Atmospheric and Solar-Terrestrial Physics, 73, 1096
\bibitem[Zhukov \& Auch$\grave{e}$re(2004)]{zhu04} Zhukov, A.~N., \& Auch$\grave{e}$re, F. 2004, \aap, 427, 705

\end{thebibliography}
\end{document}